\documentclass[aps,prl,twocolumn,showpacs]{revtex4}

\usepackage{graphicx}
\usepackage{dcolumn}
\usepackage{bm}
\usepackage[dvips]{epsfig}
\usepackage{natbib}
\usepackage{array}
\usepackage{inputenc}
\usepackage{xcolor}
\usepackage{multirow}
\usepackage{amsfonts}
\usepackage{amssymb}

\begin{document}

\title{Random walk of a swimmer in low-Reynolds-number conditions}
\author{Micha\"{e}l Garcia, Stefano Berti,  Philippe Peyla,  Salima Rafa\"\i}
\affiliation{ Univ. Grenoble 1 / CNRS, LIPhy UMR 5588, Grenoble, F-38041, France}
\begin{abstract}
Swimming at a micrometer scale demands particular strategies. When inertia is negligible compared to viscous forces,
hydrodynamics equations are reversible in time. To achieve
propulsion, microswimmers must therefore deform in a
way that is not invariant under time reversal. Here, we investigate
dispersal properties of the
micro-alga \textit{ Chlamydomonas Reinhardtii}, by means of
microscopy and cell tracking. We show that tracked trajectories
are well modeled by a correlated random walk. This process is based
on short time correlations in the direction of movement called
persistence. At longer times, correlation is lost and a standard
random walk characterizes the trajectories. Moreover, high speed
imaging enables us to show how the back-and-forth motion of flagella at very short
times affects the statistical description of the dynamics. Finally we
show how drag forces modify the characteristics of this particular
random walk.
\end{abstract}


 \maketitle

Cell motility~\cite{bray2001cell} is crucial to many biological
processes including reproduction, embryogenesis, infection, etc. Many microorganisms are able
to propel themselves, bacteria, sperm cells, microalgae, etc. A quantitative understanding of the hydrodynamics of flagella and cilia is thus of great interest~\cite{ramaswamy2010,saintillan_viewpoint}.

 One of the peculiarities of the swimming of microorganisms is that it occurs at very low Reynolds numbers which is very different from our
usual experience of swimming at our meter length
scale~\cite{Purcell1977, Taylor1951}. Indeed when inertia is
negligible as compared to viscous forces (\textit{i.e.} Reynolds
number $Re$ is lower than unity), in order to achieve propulsion,
swimmers must deform in a way that is not invariant under time reversal. This is known as Purcell's scallop theorem~\cite{Purcell1977}. In living
systems, several different strategies are used to achieve propulsion in such conditions: the \textit{E. coli} bacterium  uses a rotating flagellum at the ''back'' of its body, sperm cell propulsion relies on the asymmetry of their flagellar bending waves,  the power and recovery strokes of the two front flagella of \textit{Chlamydomonas Reinhardtii} (\textit{CR}) algae are asymmetrical.

Flagellar propulsion in \textit{CR} induces complex
swimming behavior of cells. Over short time scales, the cells undergo an
oscillating movement with changes in velocity direction
occurring at the same frequency as the beating frequency of
flagella. On a time scale longer than the period of beating, average swimming behavior is directional. Eventually, on larger
time scales, direction is lost and swimming trajectories resemble a
random walk.

\textit{CR} is a 10$\mu$m motile bi-flagellated unicellular alga.  The cell is spheroidal in shape
with two anterior flagella~\cite{sourcebook}. It belongs to the puller type of
swimmers as it uses its front flagella to propel itself, producing what a breaststroke-like movement.  The swimming direction of the cells
can be controlled by stimulus gradients, a phenomenon known as taxis,
such as chemotaxis, rheotaxis or phototaxis. Gradients are not used
in our experiments in order to avoid any external tropism on the
motility. Wild-type strains were obtained from the IBPC lab in Paris
\cite{sandrine}. Synchronous cultures of CR were grown in a
Tris-Acetate Phosphate medium (TAP) using a $12/12$ hour light/dark
cycle at $22^{o}$C. Cultures were typically grown for two days
under fluorescent lighting before the cells were harvested for
experiments.

\begin{figure}
  \begin{tabular}{ll}
\raisebox{4cm}{a)}&  \includegraphics[angle=0, height=4.5cm]{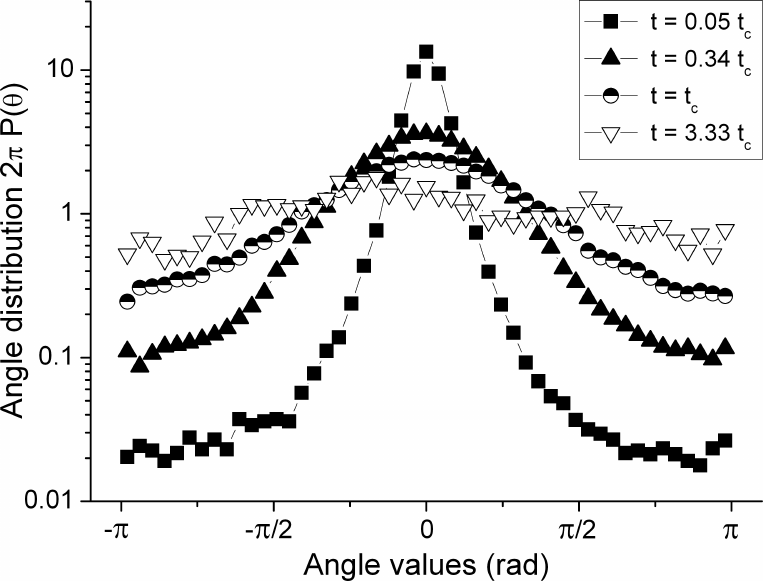} \\
\raisebox{4cm}{b)} &   \includegraphics[angle=0,height=4.5cm]{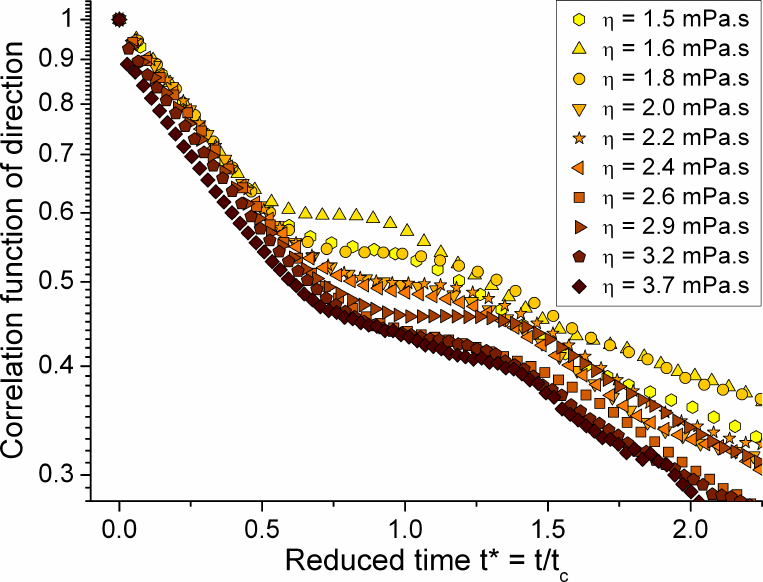}\\
 \end{tabular}
  \caption{(color online)  (a) Probability distribution functions
of angles $\theta= \arccos[\hat{k}(t_0).\hat{k}(t_0+t)]$ for
different times $t$ ranging from $0.05t_c$ to $3.33t_c$. Here
$t_c$=2.3s and the viscosity of the medium is $2.4$mPa s. Only few
data points are displayed for the sake of clarity. (b) Correlation
functions of direction $C(t/t_{c})$ as defined in the text (Log-Lin scale). Time
has been rescaled by the decaying time of an exponential. The different
symbols correspond to the viscosities used by varying the concentration of dextran.}
  \label{corr_long}
\end{figure}

   We studied the swimming  dynamics of this microorganism by means of bright field microscopy imaging on an Olympus inverted microscope coupled either
to a CCD camera (Sensicam, Photon Lines) used at a frame rate of 10Hz or to a high speed CCD camera (Miro, Phantom) used at a frame rate
of 400Hz. Long time experiments used a $\times 10$ magnification lens whereas we used a $\times 64$ lens for high speed imaging. Two hundred micrometer thick glass chambers were coated with bovine serum albumine to prevent cell adhesion. The imaged cells were located 30 to 60 micrometers from the glass walls. A red light filter was used in order to prevent phototaxis. Cell tracking~\cite{crocker} was performed using IDL (Interactive Data
Language) with a submicron precision in the detection of hundreds of cells (high speed experiments) and thousands of cells for long time sequences. To quantify the effect of drag on the cell dynamics, small amounts of short chain dextran  (Sigma Aldrich) were added to the
culture medium. The chains were short enough for non-Newtonian
effects to be absent and long enough to avoid damaging the cells
with osmotic effects. This allowed the viscosity $\eta$ of
the medium to be varied between $1.5$ and $3.7$mPa s. The range of viscosity is
restricted to this interval to ensure the viability of the cells.

Let us first recall the global dynamics of swimming, i.e. over
time scales of the order of a few seconds. Cell trajectories are
found to be correctly modelled by a persistent random
walk~\cite{randommodels, Patlak1953,golestanian2007}. Cells swim in an
almost fixed direction for a typical time of about one second. This stage corresponds to a ballistic regime characterized
by a mean velocity $V$. The ballistic regime ends when the swimmers make a turn. A new direction of motion is then observed due
to the desynchronization of the pair of
flagella~\cite{Polingoldstein2009}. At long time scales, the dispersal
properties of the swimmers are random-like \cite{goldstein2009,rafai2010}. To describe this specific random walk quantitatively, we measured
different statistical quantities of interest. Let us first define a
persistence angle $\theta (t) = \arccos
[\hat{k}(t_0).\hat{k}(t_0+t)]$, where $\hat{k}(t)$ is a unitary
vector in the direction of movement at time $t$. Thus, a value of $\theta$ close to zero reflects a certain persistence
of the trajectory. We measured the probability distribution function
of angles $\theta$ for different times $t$. At short time scales, angle
distribution peaks at around zero, characterizing the directional persistence in swimming trajectories. Over longer times, we observed a
broadening of the distribution and eventually we ended up with an
equidistribution of angle values characteristic of a random walk (figure~\ref{corr_long}(a).
This phenomenon is even better quantified by looking at the mean
value of angle distribution or equivalently at the correlation
function of direction defined as $C(t) = <\hat{k}(t_0).\hat{k}(t_0+t)>$,
where  $<>$ is an average over time $t_0$ and over all
tracked trajectories. Correlations with infinite decay time
($C(t)=1$ for all $t>0$) correspond to direction correlations preserved over
arbitrarily long times, i.e. a purely ballistic regime; whereas a
zero life-time ($C(t)=0$ for all $t>0$) corresponds to standard random walk
behavior (figure~\ref{corr_long}(b). The correlation functions decay exponentially
over a characteristic time $t_c$. This correlation time
$t_c$ is related to the mean time of persistence over which the direction
of swimming is preserved. The different symbols in
figure~\ref{corr_long}.b correspond to experiments where the
concentration of short chain dextran was varied, hence modifying
the viscosity of the medium from 1.5mPa s to 3.7mPa s. As viscosity increases, correlation time $t_c$ increases (data not
shown) from 1.5 to 3.9 seconds.

The global dynamics of swimming of \textit{CR} can
thus be described as a correlated random walk characterized by a
ballistic regime (with a mean velocity $V$) and a
decorrelation process (over a characteristic time $t_c$) due to the
turns made by the cells. As a consequence, a persistence length
$\mathcal{L}$ is naturally defined as the product $V t_c$.
From a statistical point of view, such a bahaviour is described by the mean square displacement of
cells $<r^2(t)>$ which is linear for long times ($t \gg t_c$) and
quadratic at shorter times ($t\lesssim t_c$) \cite{goldstein2009,rafai2010}. At even shorter times, the dynamics reflect the consequences of low Reynolds swimming, i.e. a non-reciprocal movement of flagella. This then leads to a zigzagging motion of cells due to the back-and-forth motion of flagella~\cite{RUFFER1985}.

\begin{figure}
\includegraphics[angle=0, width=0.8\columnwidth]{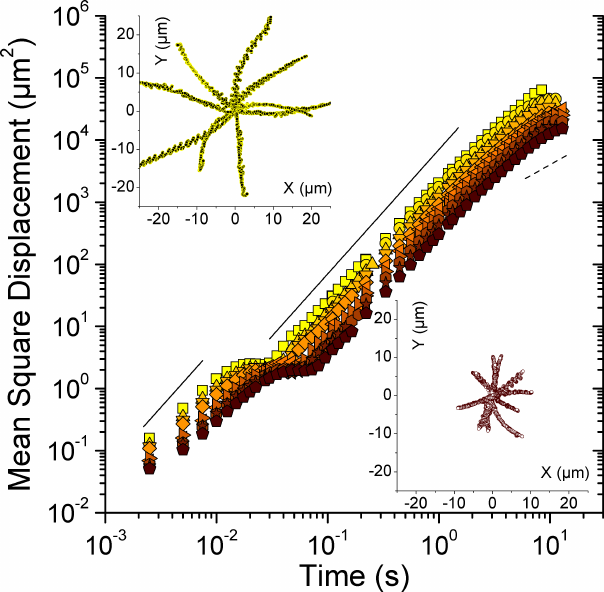}
\caption{ (color online) Mean Square Displacements $<r^{2}(t)>$ of cells as function of time for different viscosities of the medium. The different symbols represent different bath viscosities. The legend for symbols is the same as in figure~\ref{corr_short}(b). The solid lines represent a slope 2 and the dotted line a slope 1 in a log-log scale. Typical 2D
trajectories of a few cells are displayed in the insets.  In the top left inset, cells are swimming in
the nutritive medium (viscosity $\eta=1.5$mPa s), in the lower right
corner, the medium is rich in dextran ($\eta=3.7$mPa s).
Trajectories are represented on the same scale for better
comparison and both lasted 0.5 s, their starting positions were all shifted to the origin.}
    \label{msd}
\end{figure}

\begin{figure}
    \begin{tabular}{ll}
      \raisebox{4cm}{a)}  &  \includegraphics[angle=0,height=4.5cm]{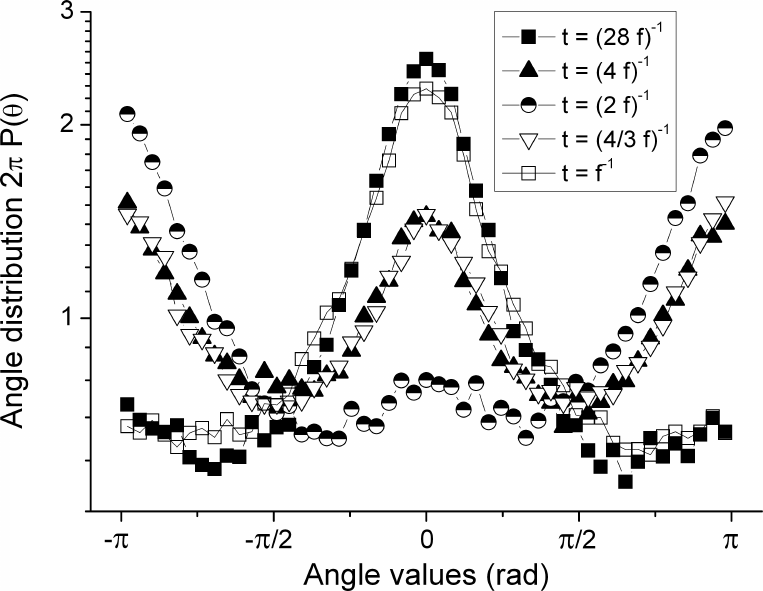} \\
      \raisebox{4cm}{b)}  &   \includegraphics[angle=0,height=4.5cm]{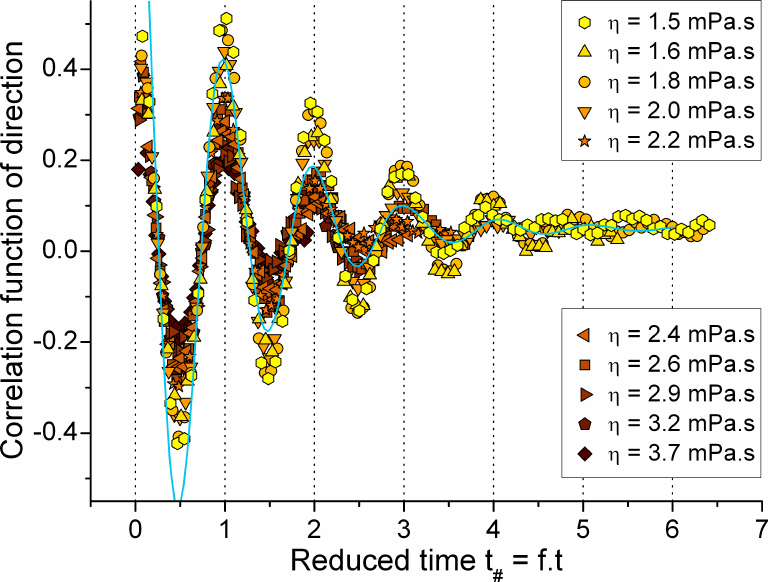}\\
    \end{tabular}
\caption{ (color online)  (a) Probability distribution functions of angle $\theta$ at
short time scales ranging from $t=1/(28f)$ to $t=1/f$. In these
experiments, cells are swimming in a medium of viscosity
$\eta=2.4$mPa s so that their beating frequency is $f =14.3$Hz. (b) Correlation functions of direction as defined in
the text. Time has been rescaled with the period $1/f$ of the signal which corresponds to the beating frequency of flagella. Symbols correspond to different viscosities of the
medium. The solid
line represents the function $\cos(2\pi t_{\#})\exp(-t_{\#})$.}
  \label{corr_short}
\end{figure}

In the present case, cells of diameter $2R\sim10\mu$m are moving at a velocity V around $50\mu$m/s in a water like medium (viscosity $\eta \sim
1$mPa s and density $\rho \sim 10^{3}$kg/m$^{3}$). This represents a
very low Reynolds number of the order of $Re = \rho V R/ \eta  \sim 2.5\times10^{-4}$. The
propulsion strategy of \textit{CR} consists in swimming
in a kind of breaststroke where the pair of flagella are wide open
during the forward movement and folded along the cell
body during the backward movement~\cite{ringo1967}. Hence, viscous
friction is high when the pair of flagella are fully extended
during the forward movement and friction is lower during the
backward movement. The symmetry under time reversal is thus broken and propulsion is ensured. However, because inertia has
no role in this regime, this kind of propulsion leads to a back-and-forth movement of the cell in which the velocity is alternatively
positive and negative. High speed imaging (400Hz) allows us to
resolve the very short time dynamics due to flagella beating and
thus to study the consequences of this back-and-forth movement on
the properties of the swimmers' random walk.

The insets in figure~\ref{msd} show typical cell trajectories imaged
at 400Hz: the back-and-forth movement of swimmers due to
the absence of inertia ($Re<<1$) together with the long time
swimming behavior are visible. In these examples, the cells are swimming either in a nutritive medium of viscosity $\eta=1$mPa s (top left inset) or
in a dextran-rich medium of viscosity $3.7$mPa s (bottom right inset). 
Cells have a net forward movement corresponding to the power stroke, followed by the
recovery stroke that propels the cell backward. As the distance
traveled forward is longer than the backward movement, the cells
ultimately progress forward. However, these fluctuations in the
direction of the velocity have consequences on the measured
statistical quantities~\cite{peruani2007} that we will discuss now.

The measured mean square displacement $<r^{2}(t)>$ shows a plateau region at very
short time ($t \ll t_c$) that reflects the transition between two
quadratic regimes : on the one hand, a fast ballistic
regime characterized by the instantaneous velocity $u$ of swimmers
and, on the other hand, a slower ballistic regime corresponding to
the mean velocity $V$ of swimming which is the resulting
forward velocity over several back-and-forth movements. The position of the
plateau therefore corresponds to the beating frequency $f$ of the swimmer,
which depends on the viscosity of the surrounding medium. To
quantify the back-and-forth swimming motion of the cells, we measured the angle
probability distribution function.
Figure~\ref{corr_short}.a shows distribution functions for
different times. For a given short time $t$, the distribution of angles $\theta (t)$ as defined earlier, peaks at around zero, reflecting a given direction at very short time.
For longer time scales (close to 1/2f), anti-correlation in cell direction resulted in new distribution peaks at values of around $\pm \pi$. When a
new stroke is produced, the measured angle is again close to
zero giving rise to a peak around zero. Hence, angle distributions
have a periodicity which reflects the beating frequency. This is
shown in figure~\ref{corr_short}.a as the
distributions are very similar at times shifted by $1/(2f)$, where
the typical frequency of the beating $f$ is deduced from the
periodical nature of the correlation function of direction.
Figure~\ref{corr_short}.b shows such correlation functions at varying $f \times t$, the product of time multiplied by the fitted frequency of
the signal. Data are well described by an exponentially attenuated
cosine function. The different symbols correspond to different viscosities of the medium. The exponential decay of the correlation function should reflect the turns in direction the cells eventually perform within a characteristic time $t_{c}$. However, due to the 3D nature of the trajectories and the 2D geometry of our setup, correlation was attenuated faster than that. 

\begin{figure}
    \begin{tabular}{ll}
      \raisebox{4cm}{a)}  &  \includegraphics[angle=0,height=4.5cm]{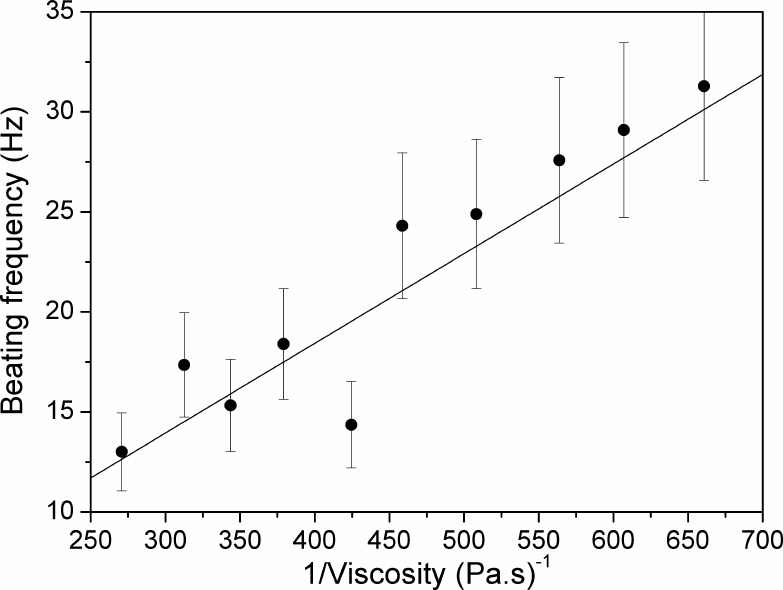} \\
      \raisebox{4cm}{b)}  &   \includegraphics[angle=0,height=4.5cm]{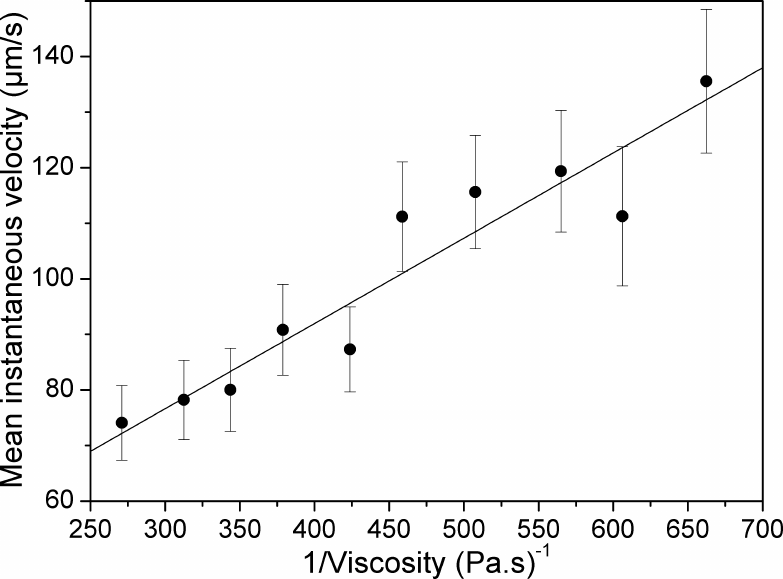}\\
    \end{tabular}
\caption{(a) Beating frequency, obtained from the period of oscillations in the correlation functions of directions, versus the inverse viscosity of the medium. (b) mean modulus of velocity $u$ as a function of the inverse viscosity of the medium. Solid lines represent linear regression giving:  $\eta f = 0.045 \pm 0.01$Pa and $\eta u = 0.15 \pm 0.04$Pa.$\mu$m }
  \label{inveta}
\end{figure}

The other consequence of the fact that swimming is produced at low
Reynolds number is that propulsion requires non-zero drag
forces. Viscous friction is thus crucial in the dynamics of
microswimmers. By varying the viscosity of the medium, we were able
to draw some conclusions about the effects of friction forces on the
locomotion of microorganisms such as this micro-alga. 

Here, short time dynamics of swimming can be fully described by few mean
quantities: flagella frequency beating $f$ (deduced from a cosine fit in figure~\ref{corr_short}.b) and the  mean modulus of instantaneous velocity $u$, which is the velocity achieved during a power or a recovery stroke. We studied the effects of viscous forces on these quantities. Velocities and beating frequency are found to be inversely proportional to the viscosity of the bath (figure~\ref{inveta}). As viscosity increases, the beating frequency decreases, varying from 30Hz to 13 Hz with a viscosity variation from 1.5 to 3.7mPa s (figure~\ref{inveta}(a) giving a slope $\eta f = 0.045 \pm 0.01$Pa. Accordingly, velocity decreases from 135 to 75$\mu$m/s (figure~\ref{inveta}(b) giving a slope $\eta u = 0.15 \pm 0.04$Pa $\mu$m. These results supports the idea of imposed-force locomotion \cite{rafai2010}. The corresponding stall force, which is proportional to the product $\eta \times u$, is then constant.

 The velocity $u$ can be related to the mean propulsion force on the cell body by Stokes' law. Let's now assume that a power stroke (respectively a recovery stroke) results from the friction length $\xi_{\perp}$ (resp. $\xi_{\parallel}$) of the flagella moving perpendicular (resp. parallel) to its long axis. Moreover, the beating frequency $f$ can be related to the friction of flagella acting on a typical distance of one cell diameter: $6 \pi \eta R u = 2R f \eta (\xi_{\perp}+\xi_{\parallel})$. Using measurements of the slopes $\eta u$ and $\eta f$, we can estimate a sum $\xi_{\perp}+\xi_{\parallel} \sim 32\mu m$. Using the friction coefficient expressions of a cylinder given in \cite{tirado1984}, this leads to an aspect ratio of 200 for a 10$\mu m$ long flagellum with a radius of 25nm. This is a reasonable estimate~\cite{sourcebook} considering that flagella are not exactly perpendicular and parallel to the flow during power and recovery strokes.

In this work, we quantified the complex dynamics of swimming at short time scale using high speed microscopy imaging and particle tracking techniques. This study allowed us to analyze the breaststroke like swimming of a CR cell in a fluid at low Reynolds number and how this swimming is influenced by the viscosity of the ambient fluid. It showed how the friction acting on a CR cell can be qualitatively extracted from the back-and-forth motion of a thin and elongated pair of flagella. This means that a description in terms of time averaged flows is not to be encouraged for such systems \cite{saintillan_viewpoint,Gollub2010_oscillatoryflow}. 

\begin{acknowledgments}
This work was financed by the Rhone-Alpes
Region (Cible program). We also acknowledge ANR Mosicob for its partial support.
 \end{acknowledgments}


\begin{thebibliography}{19}
\expandafter\ifx\csname natexlab\endcsname\relax\def\natexlab#1{#1}\fi
\expandafter\ifx\csname bibnamefont\endcsname\relax
  \def\bibnamefont#1{#1}\fi
\expandafter\ifx\csname bibfnamefont\endcsname\relax
  \def\bibfnamefont#1{#1}\fi
\expandafter\ifx\csname citenamefont\endcsname\relax
  \def\citenamefont#1{#1}\fi
\expandafter\ifx\csname url\endcsname\relax
  \def\url#1{\texttt{#1}}\fi
\expandafter\ifx\csname urlprefix\endcsname\relax\def\urlprefix{URL }\fi
\providecommand{\bibinfo}[2]{#2}
\providecommand{\eprint}[2][]{\url{#2}}

\bibitem[{\citenamefont{Bray}(2001)}]{bray2001cell}
\bibinfo{author}{\bibfnamefont{D.}~\bibnamefont{Bray}},
  \emph{\bibinfo{title}{{Cell movements: from molecules to motility}}}
  (\bibinfo{publisher}{Routledge}, \bibinfo{year}{2001}).

\bibitem[{\citenamefont{Ramaswamy}(2010)}]{ramaswamy2010}
\bibinfo{author}{\bibfnamefont{S.}~\bibnamefont{Ramaswamy}},
  \bibinfo{journal}{Annual Review of Condensed Matter Physics}
  \textbf{\bibinfo{volume}{1}} (\bibinfo{year}{2010}).

\bibitem[{\citenamefont{Saintillan}(2010)}]{saintillan_viewpoint}
\bibinfo{author}{\bibfnamefont{D.}~\bibnamefont{Saintillan}},
  \bibinfo{journal}{Physics} \textbf{\bibinfo{volume}{3}}, \bibinfo{eid}{84}
  (\bibinfo{year}{2010}).

\bibitem[{\citenamefont{Purcell}(1977)}]{Purcell1977}
\bibinfo{author}{\bibfnamefont{E.~M.} \bibnamefont{Purcell}},
  \bibinfo{journal}{Am. J. Phys.} \textbf{\bibinfo{volume}{45}},
  \bibinfo{pages}{3} (\bibinfo{year}{1977}).

\bibitem[{\citenamefont{Taylor}(1951)}]{Taylor1951}
\bibinfo{author}{\bibfnamefont{G.}~\bibnamefont{Taylor}},
  \bibinfo{journal}{Proceedings of the Royal Society of London. Series A.
  Mathematical and Physical Sciences} \textbf{\bibinfo{volume}{209}},
  \bibinfo{pages}{447} (\bibinfo{year}{1951}).

\bibitem[{\citenamefont{David~Stern}(2008)}]{sourcebook}
\bibinfo{editor}{\bibfnamefont{G.~W.} \bibnamefont{David~Stern},
  \bibfnamefont{Elizabeth~Harris}}, ed., \emph{\bibinfo{title}{The
  Chlamydomonas Sourcebook}} (\bibinfo{publisher}{Academic},
  \bibinfo{year}{2008}).

\bibitem[{san()}]{sandrine}
\bibinfo{note}{Physiologie Membranaire et Mol{\'e}culaire du Chloroplaste, UMR
  7141, CNRS et Universit\'e Paris VI}.

\bibitem[{\citenamefont{Crocker and Grier}(1996)}]{crocker}
\bibinfo{author}{\bibfnamefont{J.}~\bibnamefont{Crocker}} \bibnamefont{and}
  \bibinfo{author}{\bibfnamefont{D.}~\bibnamefont{Grier}},
  \bibinfo{journal}{Journal of Colloid and Interface Science}
  \textbf{\bibinfo{volume}{179}}, \bibinfo{pages}{298} (\bibinfo{year}{1996}).

\bibitem[{\citenamefont{Codling et~al.}(2008)\citenamefont{Codling, Plank, and
  Benhamou}}]{randommodels}
\bibinfo{author}{\bibfnamefont{E.}~\bibnamefont{Codling}},
  \bibinfo{author}{\bibfnamefont{M.}~\bibnamefont{Plank}}, \bibnamefont{and}
  \bibinfo{author}{\bibfnamefont{S.}~\bibnamefont{Benhamou}},
  \bibinfo{journal}{Journal of the Royal Society Interface}
  \textbf{\bibinfo{volume}{5}}, \bibinfo{pages}{813} (\bibinfo{year}{2008}).

\bibitem[{\citenamefont{Patlak}(1953)}]{Patlak1953}
\bibinfo{author}{\bibfnamefont{C.}~\bibnamefont{Patlak}},
  \bibinfo{journal}{Bulletin of Mathematical Biology}
  \textbf{\bibinfo{volume}{15}}, \bibinfo{pages}{311} (\bibinfo{year}{1953}).

\bibitem[{\citenamefont{Howse et~al.}(2007)\citenamefont{Howse, Jones, Ryan,
  Gough, Vafabakhsh, and Golestanian}}]{golestanian2007}
\bibinfo{author}{\bibfnamefont{J.}~\bibnamefont{Howse}},
  \bibinfo{author}{\bibfnamefont{R.}~\bibnamefont{Jones}},
  \bibinfo{author}{\bibfnamefont{A.}~\bibnamefont{Ryan}},
  \bibinfo{author}{\bibfnamefont{T.}~\bibnamefont{Gough}},
  \bibinfo{author}{\bibfnamefont{R.}~\bibnamefont{Vafabakhsh}},
  \bibnamefont{and}
  \bibinfo{author}{\bibfnamefont{R.}~\bibnamefont{Golestanian}},
  \bibinfo{journal}{Physical Review Letters} \textbf{\bibinfo{volume}{99}},
  \bibinfo{pages}{48102} (\bibinfo{year}{2007}).

\bibitem[{\citenamefont{Polin et~al.}(2009)\citenamefont{Polin, Tuval,
  Drescher, Gollub, and Goldstein}}]{Polingoldstein2009}
\bibinfo{author}{\bibfnamefont{M.}~\bibnamefont{Polin}},
  \bibinfo{author}{\bibfnamefont{I.}~\bibnamefont{Tuval}},
  \bibinfo{author}{\bibfnamefont{K.}~\bibnamefont{Drescher}},
  \bibinfo{author}{\bibfnamefont{J.~P.} \bibnamefont{Gollub}},
  \bibnamefont{and} \bibinfo{author}{\bibfnamefont{R.~E.}
  \bibnamefont{Goldstein}}, \bibinfo{journal}{Science}
  \textbf{\bibinfo{volume}{325}}, \bibinfo{pages}{487} (\bibinfo{year}{2009}).

\bibitem[{\citenamefont{Leptos et~al.}(2009)\citenamefont{Leptos, Guasto,
  Gollub, Pesci, and Goldstein}}]{goldstein2009}
\bibinfo{author}{\bibfnamefont{K.}~\bibnamefont{Leptos}},
  \bibinfo{author}{\bibfnamefont{J.}~\bibnamefont{Guasto}},
  \bibinfo{author}{\bibfnamefont{J.}~\bibnamefont{Gollub}},
  \bibinfo{author}{\bibfnamefont{A.}~\bibnamefont{Pesci}}, \bibnamefont{and}
  \bibinfo{author}{\bibfnamefont{R.}~\bibnamefont{Goldstein}},
  \bibinfo{journal}{Physical Review Letters} \textbf{\bibinfo{volume}{103}},
  \bibinfo{pages}{198103} (\bibinfo{year}{2009}).

\bibitem[{\citenamefont{Rafai et~al.}(2010)\citenamefont{Rafai, Jibuti, and
  Peyla}}]{rafai2010}
\bibinfo{author}{\bibfnamefont{S.}~\bibnamefont{Rafai}},
  \bibinfo{author}{\bibfnamefont{L.}~\bibnamefont{Jibuti}}, \bibnamefont{and}
  \bibinfo{author}{\bibfnamefont{P.}~\bibnamefont{Peyla}},
  \bibinfo{journal}{Physical Review Letters}
  \textbf{\bibinfo{volume}{104}}, \bibinfo{pages}{098102}
  (\bibinfo{year}{2010}).

\bibitem[{\citenamefont{U.~R{\"u}ffer}(1985)}]{RUFFER1985}
\bibinfo{author}{\bibfnamefont{W.~N.} \bibnamefont{U.~R{\"u}ffer}},
  \bibinfo{journal}{Cell Motility and the Cytoskeleton}
  \textbf{\bibinfo{volume}{5}}, \bibinfo{pages}{251} (\bibinfo{year}{1985}).

\bibitem[{\citenamefont{Ringo}(1967)}]{ringo1967}
\bibinfo{author}{\bibfnamefont{D.}~\bibnamefont{Ringo}}, \bibinfo{journal}{The
  Journal of Cell Biology} \textbf{\bibinfo{volume}{33}}, \bibinfo{pages}{543}
  (\bibinfo{year}{1967}).

\bibitem[{\citenamefont{Peruani and Morelli}(2007)}]{peruani2007}
\bibinfo{author}{\bibfnamefont{F.}~\bibnamefont{Peruani}} \bibnamefont{and}
  \bibinfo{author}{\bibfnamefont{L.}~\bibnamefont{Morelli}},
  \bibinfo{journal}{Physical Review Letters} \textbf{\bibinfo{volume}{99}},
  \bibinfo{pages}{10602} (\bibinfo{year}{2007}).

\bibitem[{\citenamefont{Tirado et~al.}(1984)\citenamefont{Tirado, Martinez, and
  Delatorre}}]{tirado1984}
\bibinfo{author}{\bibfnamefont{M.}~\bibnamefont{Tirado}},
  \bibinfo{author}{\bibfnamefont{C.}~\bibnamefont{Martinez}}, \bibnamefont{and}
  \bibinfo{author}{\bibfnamefont{J.}~\bibnamefont{Delatorre}},
  \bibinfo{journal}{Journal of Chemical Physics} \textbf{\bibinfo{volume}{81}},
  \bibinfo{pages}{2047} (\bibinfo{year}{1984}).

\bibitem[{\citenamefont{Guasto et~al.}(2010)\citenamefont{Guasto, Johnson, and
  Gollub}}]{Gollub2010_oscillatoryflow}
\bibinfo{author}{\bibfnamefont{J.~S.} \bibnamefont{Guasto}},
  \bibinfo{author}{\bibfnamefont{K.~A.} \bibnamefont{Johnson}},
  \bibnamefont{and} \bibinfo{author}{\bibfnamefont{J.~P.}
  \bibnamefont{Gollub}}, \bibinfo{journal}{Physical Review Letters}
  \textbf{\bibinfo{volume}{105}}, \bibinfo{pages}{168102}
  (\bibinfo{year}{2010}).

\end{thebibliography}
\end{document}